\documentclass[conference]{IEEEtran}
\IEEEoverridecommandlockouts
\usepackage{cite}
\usepackage{amsmath,amssymb,amsfonts}
\usepackage{algorithmic}
\usepackage{graphicx}
\usepackage{textcomp}
\usepackage{xcolor}
\def\BibTeX{{\rm B\kern-.05em{\sc i\kern-.025em b}\kern-.08em
    T\kern-.1667em\lower.7ex\hbox{E}\kern-.125emX}}
\begin{document}

\title{BEANNA: A Binary-Enabled Architecture for Neural Network Acceleration \\
\thanks{}
}

\author{\IEEEauthorblockN{Caleb Terrill}
\IEEEauthorblockA{\textit{University of California, Los Angeles}\\
cterrill26@ucla.edu}
\and
\IEEEauthorblockN{Fred Chu}
\IEEEauthorblockA{\textit{University of California, Los Angeles} \\
freddychu@ucla.edu}
\and

}

\maketitle

\begin{abstract}
Modern hardware design trends have shifted towards specialized hardware acceleration for computationally intensive tasks like machine learning and computer vision. While these complex workloads can be accelerated by commercial GPUs, domain-specific hardware is far more optimal when needing to meet the stringent memory, throughput, and power constraints of mobile and embedded devices. This paper proposes and evaluates a Binary-Enabled Architecture for Neural Network Acceleration (BEANNA), a neural network hardware accelerator capable of processing both floating point and binary network layers. Through the use of a novel 16x16 systolic array based matrix multiplier with processing elements that compute both floating point and binary multiply-adds, BEANNA seamlessly switches between high precision floating point and binary neural network layers. Running at a clock speed of 100MHz, BEANNA achieves a peak throughput of 52.8 GigaOps/second when operating in high precision mode, and 820 GigaOps/second when operating in binary mode. Evaluation of BEANNA was performed by comparing a hybrid network with floating point outer layers and binary hidden layers to a network with only floating point layers. The hybrid network accelerated using BEANNA achieved a 194\% throughput increase, a 68\% memory usage decrease, and a 66\% energy consumption decrease per inference, all this at the cost of a mere 0.23\% classification accuracy decrease on the MNIST dataset.

\end{abstract}


\section{Introduction}
    Neural networks have been hugely popular in recent years due to their high applicability to a wide range of cutting edge fields including but not limited to image processing, approximation, classification, automation, and mathematics. Popular network classes such as convolutional neural networks (CNNs) and deep neural networks (DNNs) achieve exceptional performance on these tasks, but come with trade-offs like expensive computational and memory demand in CNNs, and over-parameterization and computational waste in DNNs \cite{b0}\cite{b1}. Solutions addressing these issues have included introducing sparsity to networks \cite{b2}\cite{b4}, quantizing weights and activations to use smaller bit-widths \cite{b5}\cite{b6}, and using specialized hardware such as large customized ASICs or GPUs to accelerate the calculations\cite{b7}\cite{b8}. While these powerful targeted hardware architectures are effective as accelerators to carry out intense computational demands of large networks, they are not well suited for emerging applications of AI in constrained domains such as the mobile field due to their large size, high power consumption, and massive memory requirements.

    One proposed class of networks to alleviate the traditionally high computational demand of neural networks are binarized neural networks (BNNs) \cite{b10}. With BNNs, weights and activations are constrained to either -1 or +1, significantly reducing the computation complexity of inference and lowering the needed memory to store weights. Despite these benefits, BNNs are still able to achieve impressive performance results \cite{b11}, making them a promising candidate for applications demanding a low memory footprint and reduced power consumption.
    
    Existing works on BNNs have found that in order to maintain results comparable to higher precision networks, the first and last layers must be kept at a high precision, as these layers are associated with the inputs and output of the network \cite{b9}. Meanwhile, the hidden layers are able to be binarized with a near negligible performance reduction. For a hardware designer looking to implement an accelerator for such networks, this presents a fascinating challenge of how to best design an architecture that can effectively handle both high precision and binary data types and operations. 

\subsection{Contributions}

   In this paper, we introduce a novel Binary-Enabled Architecture for Neural Network Acceleration (BEANNA), a hardware accelerator capable of processing both high precision and binary network layers. For our high precision data type, we chose bfloat16 (Brain Floating Point), with 8 exponent bits and 8 significand bits, due to its large dynamic range and simplified computational complexity compared to fp32. At the core of BEANNA is a 16x16 systolic array based matrix multiplier. When operating in binary mode, each of the 256 processing elements (PEs) of the array can compute the partial sum result of 16 binarized input activations and weights, allowing the array to effectively act as a 256x16 systolic array for processing binary layers. 

    Running at a clock speed of 100MHz on a Xilinx FPGA evaluation board, BEANNA is able to achieve a peak throughput of 52.8 GigaOps/second when operating in high precision mode, and 820 GigaOps/second when operating in binary mode. When comparing a hybrid network with binarized hidden layers to a network with only floating point layers, the hybrid network accelerated using BEANNA achieved a 194\% throughput increase, a 68\% memory usage decrease, and a 66\% energy consumption decrease per inference, all of this at the cost of a mere 0.23\% classification accuracy decrease on the MNIST dataset.
\subsection{Outline}

    The organization of this paper is as follows: Section 2 provides background on BNNs, systolic arrays, and bfloat16. Section 3 details the methods for network training and overviews the hardware architecture. Sections 4 presents and evaluates the results for BEANNA. Lastly, section 5 concludes.
\section{Background}

\subsection{Binarized Neural Networks}
Neural network computations are composed of many inner product calculations. A variety of numerical representations can be used for these computations, including floating point, half-precision floating point, 16-bit fixed point, and 8-bit fixed point. There is even existing work researching the usage of variable-bit fixed point calculations based on the dynamic range requirements of network layers \cite{b14}.

An inner product computation involving a weights vector $W$ and an input activations vector $I$, each of size $N$, requires $N$ total multiplications and $N-1$ total additions. For hardware implementations of neural networks, the complicated multiplication units required for the inner products and the large amounts of weight memory are costly. For that reason, much research has gone into methods of reducing the number of multiplications and memory usage in neural network inference. 

Binary neural networks allow for input activations or weights to be expressed using a single bit, representing either -1 or +1. In the case of both binary weights and binary activations, a multiplication is simply an XNOR, and an inner product result $s$ can be computed as:

\begin{equation}
    s = sgn(W) \oplus sgn(I)
\end{equation} 

Here, the $\oplus$ refers to an XNOR and popcount operation between vectors with single-bit elements. This simplification reduces the original multiplications and additions to simply $N$ XNORs and an $N$-bit popcount. Additionally, during inference only the bit vector $sgn(W)$ needs to be stored for weights, so the necessary weight memory decreases by a factor of the original higher precision weight bit-width. 

Courbariaux et al. showed that binarized neural networks with weights and activations constrained to -1 or +1 can achieve phenomenal performance results \cite{b9}. To train such networks, high precision weight values $W$ are binarized for inner product computations during a forward pass. For a backwards pass, the gradient of $W$ is estimated: 

\begin{equation}
    \frac{\partial L}{\partial W} \approx \frac{\partial L}{\partial (sgn(W))} = \frac{\partial L}{\partial s} \frac{\partial s}{\partial (sgn(W))} = \frac{\partial L}{\partial s} sgn(I)
\end{equation} 

The high precision $W$ values are updated with these gradients, and are clipped to fall between -1 and 1 to prevent them from growing very large without affecting the binarized weights.

\subsection{Systolic Arrays}

The key mechanism used by BEANNA to perform the intensive computations required by a neural network is a systolic array matrix multiply unit. A systolic array is a homogeneous hardware structure composed of a grid of PEs that allow for highly parallelized and efficient computations. In this system, data will flow from memory and pass through consecutive PEs that perform iterative calculations before returning the results back to memory. The structure can be 1D like a simple pipeline, 2D like a grid, or an even higher dimension--as long as there is parallelism to take advantage of. While this architecture has many applications, a major use today is to accelerate neural networks; The highly repetitive multiply-add and accumulates needed in the inference phase can be exploited by the dataflow circulation of a systolic array processor. Section 3C covers the structure and PE design of the systolic array in BEANNA.

\subsection{Brain Floating Point}
\begin{figure}[h]
    \includegraphics[width=0.47\textwidth]{"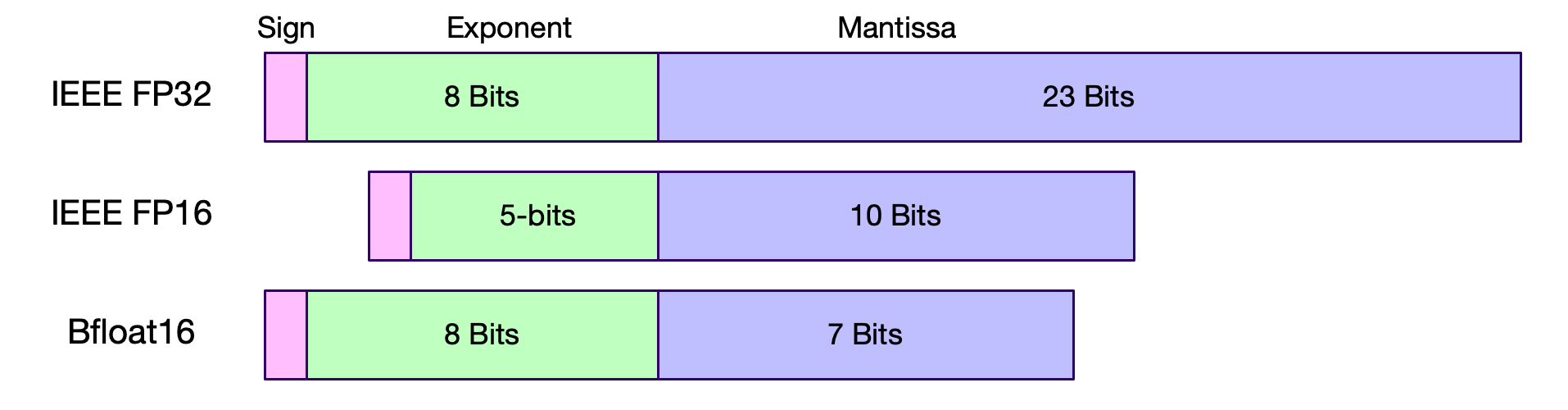"}
    \caption{Bfloat16 vs IEEE standard data types}
\end{figure} 
While neural networks can use a wide variety of data types to hold their values, some work much better than others. Google's TPU cloud accelerators \cite{b13} use a custom floating point format called bfloat16, composed of a sign bit, 8 exponent bits, and 7 mantissa bits. This  modification on the half-precision fp16 format allows for the dynamic range of fp32 at a much smaller physical multiplier size. This is since the multiplier size scales quadratically with the size of the mantissa, so using bfloat16 not only allows for a larger range than fp16, but also a smaller physical size of the hardware multiplier. Because of these benefits, the bfloat16 data type was chosen as the format for the high precision layers executed on BEANNA.

\section{Methods}
This section details the specifications of the design and the methods by which we implemented BEANNA. The neural networks were first tested and trained with PyTorch. Afterwards, the hardware architecture was coded in SystemVerilog, followed by simulation, synthesis, and implementation using Xilinx's Vivado for a Zynq Ultrascale+ MPSoC ZCU106 FPGA. 
\subsection{Training}
The neural network training for BEANNA was done on the MNIST dataset using custom PyTorch layers. The network configuration of choice was a fully-connected network of 784 input neurons, 10 output neurons, and 3 hidden layers each composed of 1024 neurons. The weights and activations of the input and output layers were floating point, while for the hybrid network the hidden layers converted their weights and activations to binary. 

For the output of each layer, a hardtanh activation function was applied, followed by a batch normalization layer.

\begin{equation}
\text{hardtanh}(x) = 
\left\{
\begin{array}{ll}
      -1 & x < -1 \\
      x & -1 \leq x \leq 1 \\
      1 & x > 1\\
\end{array}
\right.
\end{equation}

We trained both a fully floating point and a hybrid network over 100 epochs.
\begin{figure}[h]
    \includegraphics[width=0.46\textwidth]{"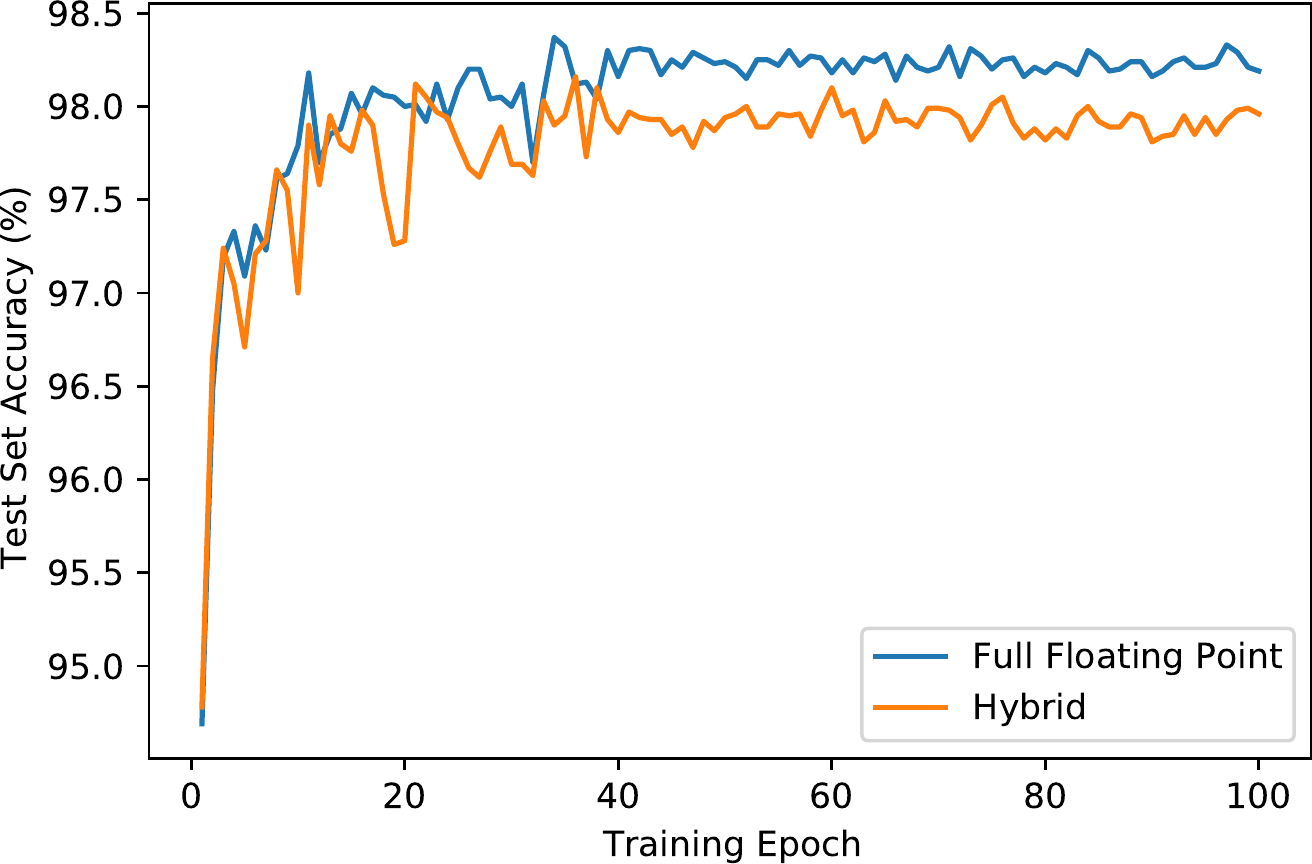"}
    \caption{Network training accuracy progression}
\end{figure} 

As shown in Figure 2, both the the fully floating point network and the hybrid binary network with the floating point edge layers improve in  accuracy  rapidly  in  the  early  epochs  of  training, and then slowly reach asymptotic max accuracies after around  50  epochs. The fully floating point network reached a final test set classification accuracy of 98.19\% and the binary network reached an accuracy of 97.96\%, a difference of 0.23\%.

\subsection{Hardware Overview}
BEANNA is a reconfigurable neural network accelerator designed for high inference throughput, a low memory footprint, and low energy usage by utilizing hidden layers of binary weights and activations. The primary mechanism by which it performs its computations is the systolic array based matrix multiply unit at its core. The top level overview of the hardware architecture is shown in Figure 3.

\begin{figure}[h]
    \includegraphics[width=0.48\textwidth]{"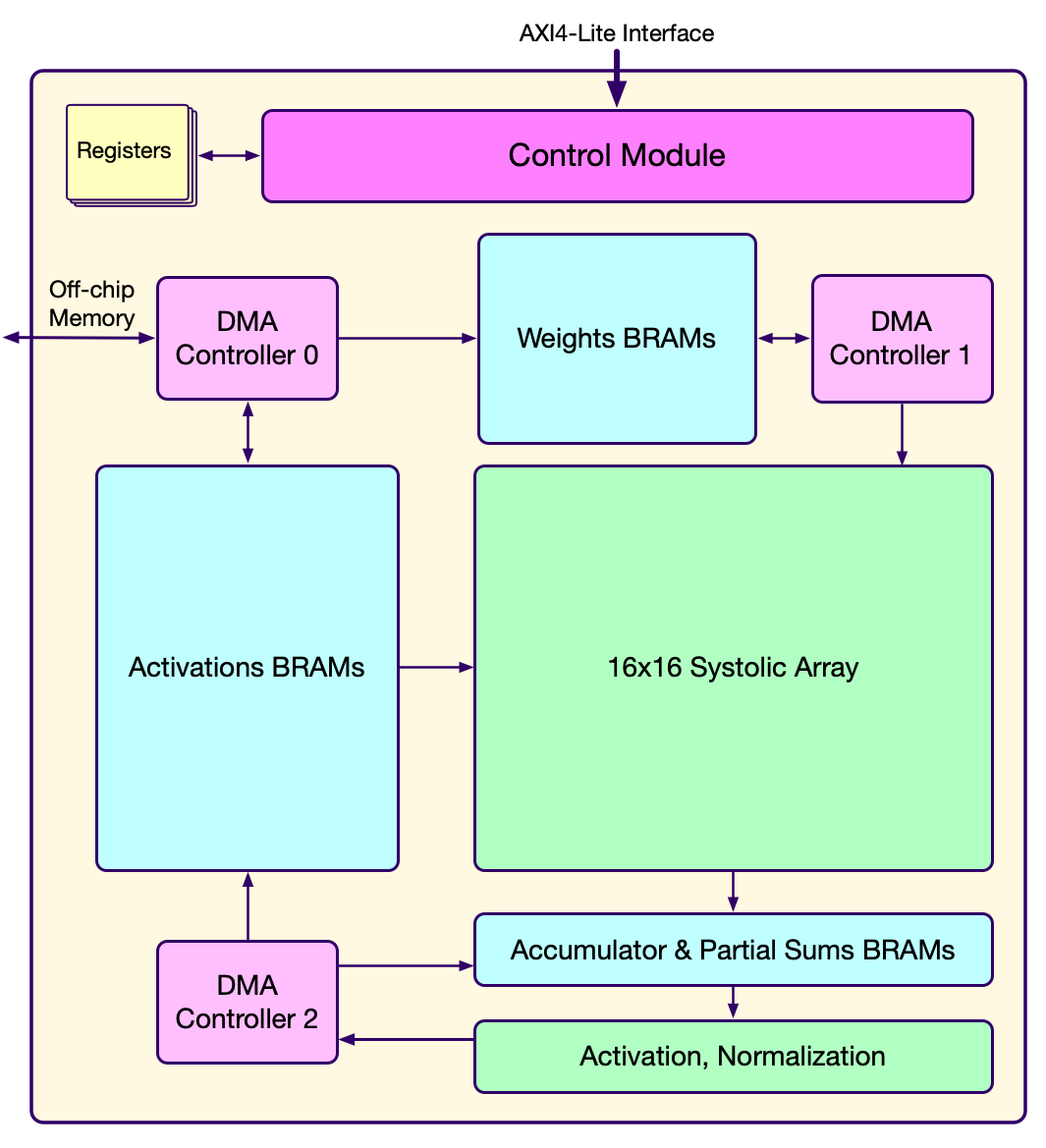"}
    \caption{High level overview of BEANNA architecture}
\end{figure} 

The architecture is composed of a main control module, Block RAMs (BRAMs) for holding activations, weights, and partial sums, direct memory access (DMA) controllers for moving data, and a systolic array based matrix multiply unit. The control module, shown at the top of Figure 3, utilizes an Advanced eXtensible Interface(AXI4)-Lite interface to communicate with software or a external hardware controller. It communicates with three DMA controllers which move the weights, activations, and partial sums throughout the design. DMA controller 0, shown in the top left of the design, is responsible for reading the trained weights and first layer activations from off-chip memory. It also writes the final results of the inference back to off-chip memory. DMA controller 1 moves the weights from the weights BRAMs into the systolic array at the beginning of each matrix multiply. DMA controller 2 reads a layer's output activations and updates the activations BRAM with the newly calculated results.

\subsection{Matrix Multiply Systolic Array}

In the systolic array for BEANNA, shown in Figure 4, the input activations flow horizontally to the right, with activations of a single batch flowing on different rows staggered by one column. The partial sums move downwards as they get accumulated in each consecutive PE until they eventually reach the partial sum accumulator BRAMs at the bottom. From there, the results are continuously accumulated as the system performs a block matrix multiplication, meaning it sums together the multiplication results of submatrices that can fit in the 16x16 systolic array. After the full multiplication is complete, the results get written back into the activations BRAM on the left after an activation function and normalization are applied. 

\begin{figure}[h]
    \includegraphics[width=0.46\textwidth]{"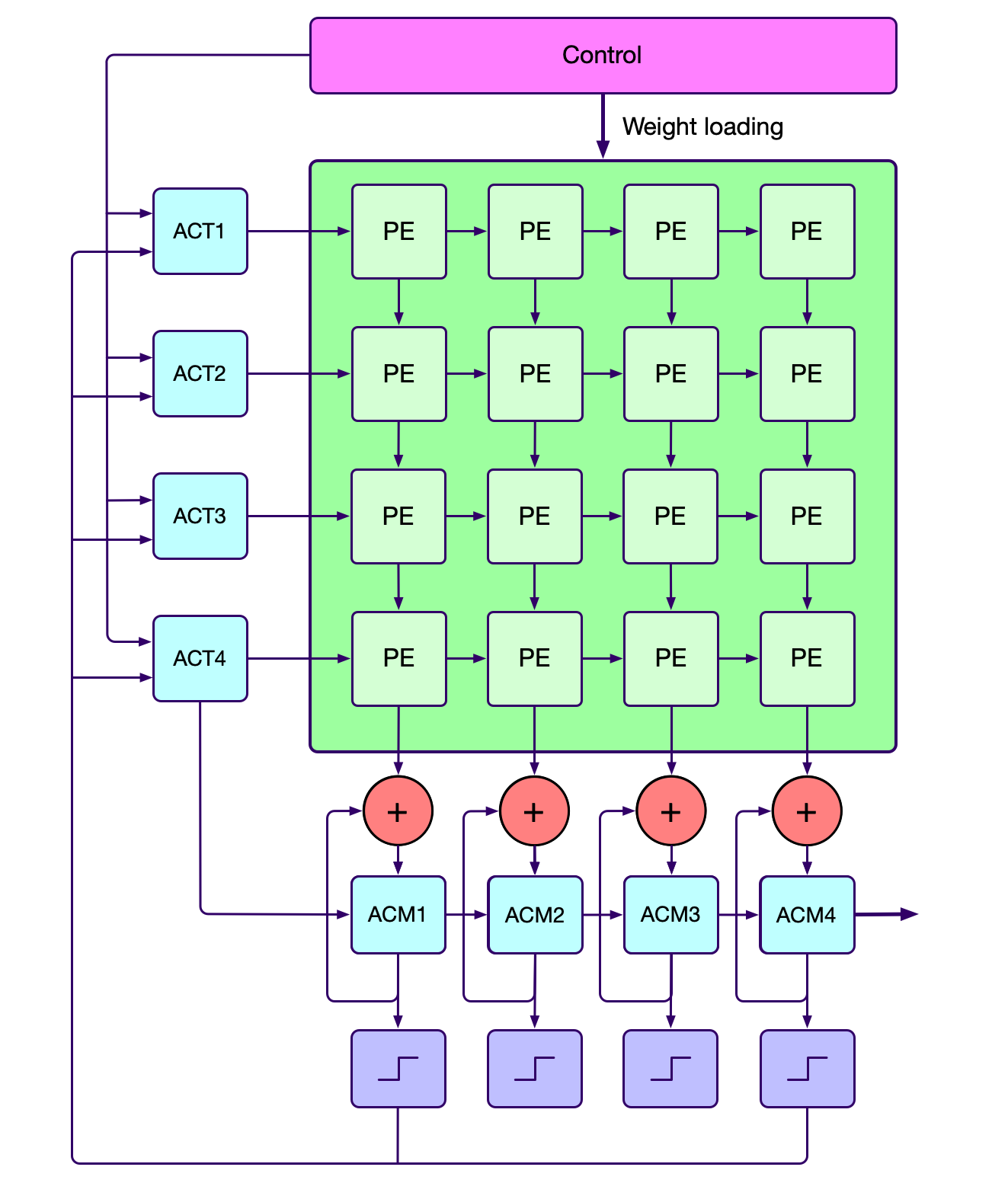"}
    \caption{Overview of matrix multiply systolic array configuration}
\end{figure} 

Figure 5. shows the design of an individual BEANNA PE. It contains two computation modules, both of which perform a multiply add on their respective data types. The multiplier for the binary module is just an element-wise 16-bit XNOR. The modules multiply together the input activation and weight values, then add the product to the input partial sum. The outputs of the computations are then muxed by a control signal, which determines the result to select depending on whether the systolic array is operating in binary mode or high precision mode. As an optimization, the control also ties off the inputs of the unused computation unit to minimize unnecessary switching power usage. The resulting partial sum is passed downwards to the PE below it, while the input activation is passed rightwards to the next PE in the row. 

\begin{figure}[h]
    \includegraphics[width=0.49\textwidth]{"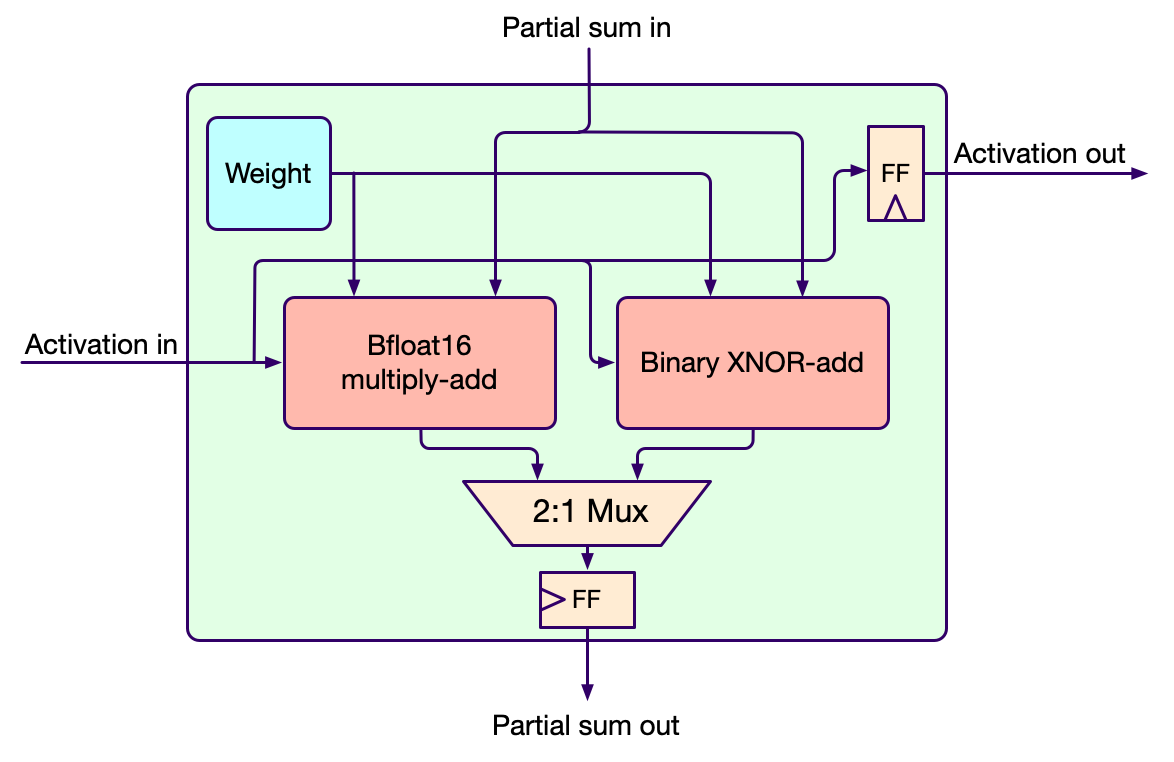"}
    \caption{Processing element (PE) in systolic array for bfloat16 and binary data types}
\end{figure}

 \subsection{Dataflow}
 The procedure for a single inference run by BEANNA is as follows. 
 \begin{enumerate}
\item Software or hardware passes instructions to a main controller through the AXI4 Lite interface.
\item DMA controller 0 reads from off-chip memory and writes activations into the activations BRAMs.
\item DMA controller 0 reads from off-chip memory and writes weights into the weights BRAM.
\item DMA controller 1 transfers weights from the weights BRAM to the systolic array.
\item The systolic array's operation mode is set to either high precision or binary, depending on the layer type.
\item Computation begins, with activations flowing to the right in the systolic array, and the partial sums flowing downwards. 
\item Values flow out from the bottom of the systolic array into the partial sums accumulators.
\item Return to 3) or 4) until the block matrix multiply is complete.
\item DMA controller 2 moves values from the partial sums accumulators through activation and normalization units, then back into the activation BRAMs.
\item Return to 3) until all the network layers are complete.
\item DMA controller 0 reads from the activations BRAMs and writes the inference results to off-chip memory.
\end{enumerate}

\section{Results}
To analyze performance, we compared a fully floating point neural network run on a hardware architecture with no binary capabilities to a hybrid neural network accelerated using BEANNA. Both networks used the same layers sizes of 784, 1024, 1024, 1024, and 10, trained on the MNIST dataset as described in Section 3A. Both networks were also run with a batch size of 1 and 256 to simulate single and batched inference.

\begin{table}[h]
\begin{center}
\caption{Performance and Speed}
\begin{tabular}{|l|l|l|l|l|}
\hline 
Parameter & Floating Point Only & BEANNA \\ \hline
Testset Accuracy            &  98.19\%    &  97.96\%     \\ \hline
Inferences/second--Batch 1            &  138.42 &  409.13 \\ \hline
Inferences/second--Batch 256  &  6928.08  &    20337.60 \\ \hline
Timing (100MHz)             &  Passed    &    Passed    \\ \hline
\end{tabular}
\end{center}
\end{table}

From Table 1 above, the hybrid network accelerated by BEANNA had much faster inference times than the floating point network by about 3x for both batch sizes of 1 and 256. This is enabled by the larger systolic array size of 256x16 when processing binary layers. The difference in performance between the two batch sizes demonstrates the strengths of the systolic array architecture, as consecutive inferences can be pipelined through the hardware, resulting in a much higher average throughput. The fastest performing configuration was the batch-size 256 hybrid network, reaching a speed of over 20,000 inferences per second. 
 
\begin{table}[h]
\begin{center}
\caption{Memory and Hardware Utilization}
\begin{tabular}{|l|l|l|l|l|l|}
\hline 
 Parameter & Floating Point Only & BEANNA \\ \hline
LUTs           &    89,838       &     102,297   \\ \hline
FFs          &    25,636       &     25,615    \\ \hline
BRAMs          &    71.5         &     71.5      \\ \hline
DSP Slices    &    256          &     256       \\ \hline
Memory Usage  &    5,820,416 bytes    &     1,888,256 bytes \\ \hline
\end{tabular}
\end{center}
\end{table}

Table 2 shows the memory and hardware utilization of a base floating point only accelerator and the BEANNA architecture. Here we see that both architectures utilize a similar amount of hardware, but with one key difference--BEANNA enables the use of binary hidden layers, and thus is able to use 3x less off-chip memory than the floating point network; This improvement stems from the inherent nature of 1-bit weights requiring significantly less storage space than 16-bit bfloat16 weights. Also note that the addition of the binary hardware for BEANNA results in only a very small increase in LUT usage, as the binary logic is fundamentally very hardware efficient. 

\begin{table}[h]
\begin{center}
\caption{Power Consumption (Batch 256)}
\begin{tabular}{|l|l|l|l|l|l|}
\hline 
\\ \hline
 Parameter & Floating Point Only & BEANNA \\ \hline
Total Power              &    2.135 W    &  2.150 W            \\ \hline
Static Power             &    0.600 W     &   0.600 W      \\ \hline
Dynamic Power            &   1.535 W     &    1.550 W   \\ \hline
Single Inference Energy         &     0.3082 mJ    &   0.1057mJ  \\ \hline
\end{tabular}
\end{center}
\end{table}

Lastly, we looked at the power results post implementation using the Vivado Power Estimator (XPE) for Optimization and Analysis. This provided us with a general idea of how much power is consumed during normal operation. The two architectures were tested by running inference on random data. Due to BEANNA's extra binary hardware, it consumed 0.015W more power than the floating point only architecture. However, since the hybrid network had much higher inference speeds, it resulted in much lower energy costs per inference. The fully floating point network required 0.3082mJ to compute a single inference, whereas the hybrid network used a third of that amount at 0.1057mJ, demonstrating BEANNA's suitability in the lower power domains like mobile devices.

It is also important to note that while Vivado and the FPGA make a fair amount of layout optimizations, designing a custom ASIC for BEANNA would result significant improvements in speed and power due to its higher degree of specificity. The next steps for BEANNA involve designing and synthesizing an ASIC to get the most high performance benchmarks for this architecture.

\section{Conclusion}
    In this paper, we propose BEANNA, a domain-specific architecture that presents a novel systolic array based matrix multiply unit that can perform both floating point and binary calculations throughout all its PEs. BEANNA is designed to be a versatile accelerator architecture that would fit the mobile and embedded systems paradigm, as it is unique in that it supports mixed layer neural network configurations of both binary and floating point layers. When a hybrid network with floating point outer layers and binary hidden layers was implemented and tested on BEANNA, it demonstrated a speedup of 3x compared to a fully floating point network, at the expense of a 0.23\% loss in accuracy on the MNIST dataset. BEANNA also outperformed the floating point network in memory and energy usage, requiring only a third of the amount of memory and a third of the total energy per inference. Overall, BEANNA--and binary neural networks in general--prove to be a promising solution to address the challenges of designing a neural network accelerator for low power, low memory, and high throughput use cases.


\begin{thebibliography}{00}
\bibitem{b0} Kaya, Y., Hong, S., \& Dumitras, T. (2019, May). Shallow-deep networks: Understanding and mitigating network overthinking. In International Conference on Machine Learning (pp. 3301-3310). PMLR.
\bibitem{b1} Gu, Y., Zhang, W., Fang, C., Lee, J. D., \& Zhang, T. (2020). How to characterize the landscape of overparameterized convolutional neural networks.
\bibitem{b2} Liu, B., Wang, M., Foroosh, H., Tappen, M., \& Pensky, M. (2015). Sparse Convolutional Neural Networks. In Proceedings of the IEEE Conference on Computer Vision and Pattern Recognition (CVPR).
\bibitem{b4} Srinivas, S., Subramanya, A., \& Venkatesh Babu, R. (2017). Training Sparse Neural Networks. In Proceedings of the IEEE Conference on Computer Vision and Pattern Recognition (CVPR) Workshops.
\bibitem{b5} Jin, Q., Yang, L., \& Liao, Z. (2020). Adabits: Neural network quantization with adaptive bit-widths. In Proceedings of the IEEE/CVF Conference on Computer Vision and Pattern Recognition (pp. 2146-2156).
\bibitem{b6} Lin, D., Talathi, S., \& Annapureddy, S. (2016, June). Fixed point quantization of deep convolutional networks. In International conference on machine learning (pp. 2849-2858). PMLR.
\bibitem{b7} Van, N. T. T., \& Thinh, T. N. (2015, November). Accelerating anomaly-based IDS using neural network on GPU. In 2015 international conference on Advanced Computing and Applications (ACOMP) (pp. 67-74). IEEE.

\bibitem{b8} Knag, P., Kim, J. K., Chen, T., \& Zhang, Z. (2015). A sparse coding neural network ASIC with on-chip learning for feature extraction and encoding. IEEE Journal of Solid-State Circuits, 50(4), 1070-1079.

\bibitem{b9} Courbariaux, M., \& Bengio, Y. (2016). BinaryNet: Training Deep Neural Networks with Weights and Activations Constrained to +1 or -1. ArXiv, abs/1602.02830.
\bibitem{b10} Saad, D.; Marom, E. Training feed forward nets with binary weights via a modified CHIR algorithm. Complex Syst. 1990, 4, 573–586. 
\bibitem{b11} Liu, Z., Shen, Z., Savvides, M., \& Cheng, K. T. (2020, August). Reactnet: Towards precise binary neural network with generalized activation functions. In European Conference on Computer Vision (pp. 143-159). Springer, Cham.
\bibitem{b12} Rastegari, M., Ordonez, V., Redmon, J., \& Farhadi, A. (2016). XNOR-Net: ImageNet Classification Using Binary Convolutional Neural Networks. In Computer Vision - ECCV 2016 - 14th European Conference, Amsterdam, The Netherlands, October 11-14, 2016, Proceedings, Part IV (pp. 525–542). Springer.

\bibitem{b13} Jouppi, Norman P. et al. "In-Datacenter Performance Analysis of a Tensor Processing Unit." Proceedings of the 44th Annual International Symposium on Computer Architecture. Association for Computing Machinery (ACM)
\bibitem{b14} N. Mitschke, M. Heizmann, K. Noffz \& R. Wittmann, "A Fixed-Point Quantization Technique for Convolutional Neural Networks Based on Weight Scaling," 2019 IEEE International Conference on Image Processing (ICIP), 2019, pp. 3836-3840
\bibitem{b15} Haotong Qin, Ruihao Gong, Xianglong Liu, Xiao Bai, Jingkuan Song, Nicu Sebe, Binary neural networks: A survey, Pattern Recognition,Volume 105, 2020, 107281,ISSN 0031-3203
\end{thebibliography}
\end{document}